\documentclass[aip,graphicx]{revtex4-1}
\usepackage{graphicx}
\usepackage{subcaption}
\usepackage{float}

\draft 
\usepackage{textcomp}
\usepackage[table]{xcolor}

\usepackage[table]{xcolor}

\definecolor{rowA}{RGB}{220,220,220} 
\definecolor{rowB}{RGB}{160,160,160} 

\begin{document}

\preprint{}

\title{Reconfigurable and cascaded logic gates using dual-input multilayered heater nanocryotrons}


\author{Behnoosh Babaghorbani}
\email[]{b.babaghorbani@tudelft.nl}
\affiliation{Department of Imaging Physics, Faculty of Applied Sciences, Delft University of Technology, 2628 CJ Delft, The Netherlands}

\author{M. Yu. Mikhailov}
\altaffiliation[Present address: ]{Single Quantum B.V., 2629 HH Delft, The Netherlands}
\affiliation{Department of Imaging Physics, Faculty of Applied Sciences, Delft University of Technology, 2628 CJ Delft, The Netherlands}
\affiliation{B.~Verkin Institute for Low Temperature Physics and Engineering of the National Academy of Sciences of Ukraine, 61103 Kharkiv, Ukraine}

\author{Hui Wang}
\affiliation{Department of Imaging Physics, Faculty of Applied Sciences, Delft University of Technology, 2628 CJ Delft, The Netherlands}

\author{Thomas Descamps}
\affiliation{Quantum Technology, Microtechnology and Nanoscience, Chalmers University of Technology, 412 96 Gothenburg, Sweden}

\author{Val Zwiller}
\affiliation{Single Quantum B.V., 2629 HH Delft, The Netherlands}

\author{Iman Esmaeil Zadeh}
\affiliation{Department of Imaging Physics, Faculty of Applied Sciences, Delft University of Technology, 2628 CJ Delft, The Netherlands}
\affiliation{Single Quantum B.V., 2629 HH Delft, The Netherlands}


\date{\today}

\begin{abstract}
Superconducting electronics have emerged as a promising platform for advanced information processing, offering unique opportunities for on chip computation and signal manipulation at cryogenic temperatures. These devices hold particular potential in applications ranging from quantum computing to high sensitivity magnetic sensing, where integrated logic and scalable circuit architectures are essential for performing complex computational and signal-processing tasks.
In this work, we present a dual-input multilayered heater nanocryotron (hTron) that introduces both multi input functionality and reconfigurable logic capability within a single device. This capability represents a step forward toward realizing more complex computational architectures. In addition, we demonstrate that these devices can, in principle, drive one another and potentially be integrated on a larger scale. Furthermore, the inherent reconfigurability of the demonstrated device allows for dynamic switching between logic operations without requiring additional components which reduces circuit area and simplifies cryogenic and biasing requirements, making the design highly suitable for scalable superconducting computing systems.
\end{abstract}


\maketitle 



 Superconducting electronics are emerging technologies for ultralow power, high speed digital circuits particularly when computation or sensing platform is naively cryogenic. The absence of resistive losses below the superconducting transition temperature and the capability to exploit superconducting phenomena allow for energy dissipation orders of magnitude lower than conventional CMOS technology\cite{Braginski2019_SuperconductorElectronics,Wang2025_AttojouleThermalSwitch}. In particular, nanowire based superconducting devices such as three terminal cryotrons (nTrons) \cite{McCaughan2014} and heater based thermal switches\cite{Wang2025_AttojouleThermalSwitch} provide a new family of building blocks for logic gates and memory, circumventing some scalability and interfacing issues found in classical Josephson Junction logic families \cite{Yasukawa2024,McCaughan2019, Likharev1991_RSFQ, Tolpygo2016_SuperconductorElectronics}. 
 Among the superconducting electronic devices introduced for on-chip operation, Single Flux Quantum (SFQ) circuits are notable for providing high-speed performance, although they suffer from low fan-out and interfacing complexities\cite{Friedman2022, Yorozu2003_SFQRouter, Pedram2019}. 

Nanocryotron (nTron) devices have enabled a compelling family of superconducting logic and memory elements
\cite{McCaughan2014, Yasukawa2024,Buzzi2023}. These devices leverage direct electrical interaction between control and channel nanowires to realize compact and efficient superconducting circuits, and have also been demonstrated as photolithography compatible three terminal switches capable of directly driving CMOS loads \cite{paul2025photolithography}.
While this approach has proven effective for dense logic and memory integration, the existence of direct electrical coupling between elements introduces constraints on bias margins, fan out, and isolation between circuit stages.
hTrons \cite{Baghdadi2020, Karam2025_hTronSPICE,Wang2025_AttojouleThermalSwitch}
enables reliable switching with complete electrical isolation. Baghdadi\cite{Baghdadi2020} et al. demonstrate an improved geometry called hTron with eliminated leakage current via electrical isolation between the heater and superconducting channel, enabling its use as a reliable switch for superconducting circuits and cryogenic electronics, including logic elements, memory interfaces, and signal routing. Subsequent studies further improved energy efficiency and switching performance through refined heater channel designs \cite{Karam2025_hTronSPICE, Wang2025_AttojouleThermalSwitch}. These specialized hTron elements highlight a complementary design approach in which leakage free electrical isolation enables scalable integration of superconducting switch networks, while switching is governed by thermal processes. 
Previous work on hTron devices has investigated their thermal and electrical behavior using SPICE-based modeling frameworks  \cite{Karam2025_hTronSPICE}, providing insight into activation delay and energy requirements.

Digital circuits built with superconducting devices offer the advantage of cryogenic integration with superconducting devices such as Superconducting Nanowire Single Photon Detectors (SNSPDs) \cite{Korzh2020_SNSPD,Colangelo2022_SNSPD_IR,Goltsman2001} and superconducting quantum computing systems\cite{Yoshikawa2019_SuperconductingDigital}, potentially eliminating the need for multiple conversion and amplification stages between the cryogenic and room temperature environments. A recently demonstrated architecture achieved logic gate operations (NOT, NAND, NOR, AND, OR) and volatile memory cells using a superconducting thermal switch with switching energies in the attojoule regime and ultralow bit error rates \cite{Wang2025_AttojouleThermalSwitch}. However, the majority of these devices remain fixed in their function, restricting system flexibility and area efficiency for scalable reconfigurable computing. 

Reconfigurable logic circuits are highly desirable in digital systems because they allow a single hardware platform to perform various logic operations depending on the control knobs, thereby reducing redundant circuit area, simplifying interconnect networks, and enabling dynamic reuse of hardware resources. Recently demonstrated cryogenic logic architectures\cite{Alam2023_hTron} have shown that 
nano-heater controlled superconducting devices  can support multiple Boolean operations; however, these functions are typically realized through analog current summation within a shared thermal region under specific bias conditions, exhibiting response times on the order of milliseconds, rather than through structurally independent and electrically isolated input pathways.

In this work, we present a dual-heater hTron that incorporates multi-input functionality and logic reconfigurability within a single nanowire-based device. By integrating two explicit hardware defined inputs as heaters instead of one, our device enables a richer set of logic functions without physically altering the layout, thereby enabling digital circuit operations that are dynamically configurable. In addition, we demonstrate how the device can drive other identical units, showing potential for cascadeable, reconfigurable logic circuits. The inherent reconfigurability simplifies cryogenic biasing and reduces interface complexity, making the design more suited to scalable superconducting computing systems. 

\begin{figure}[!htbp]
    \centering
    \begin{subfigure}{0.5\textwidth}
        \centering
        \includegraphics[width=\textwidth, keepaspectratio]{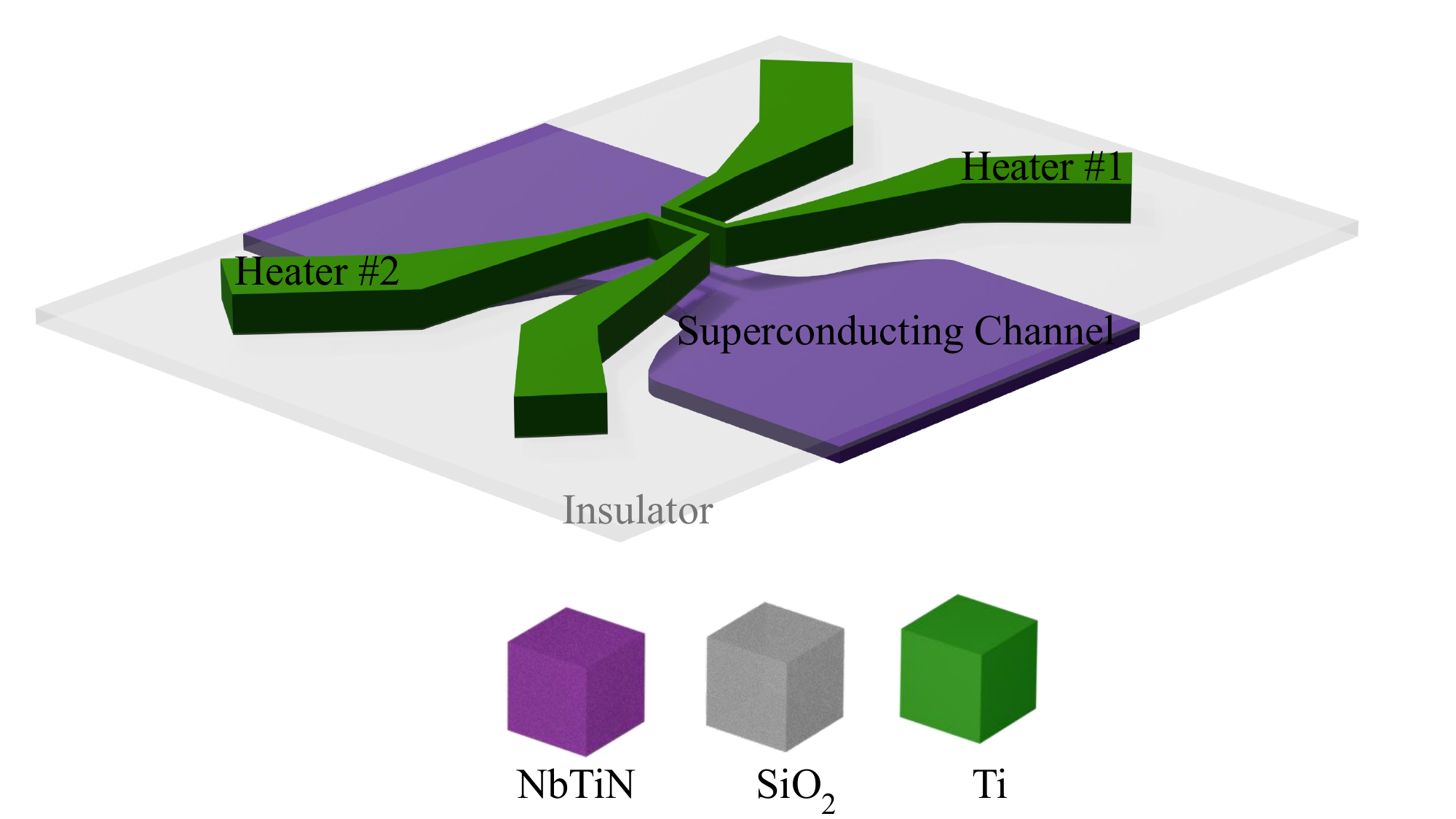}
        \caption{}
        \label{subfig:b10}
    \end{subfigure}
    \begin{subfigure}{0.4\textwidth}
        \centering
        \includegraphics[width=\textwidth, keepaspectratio]{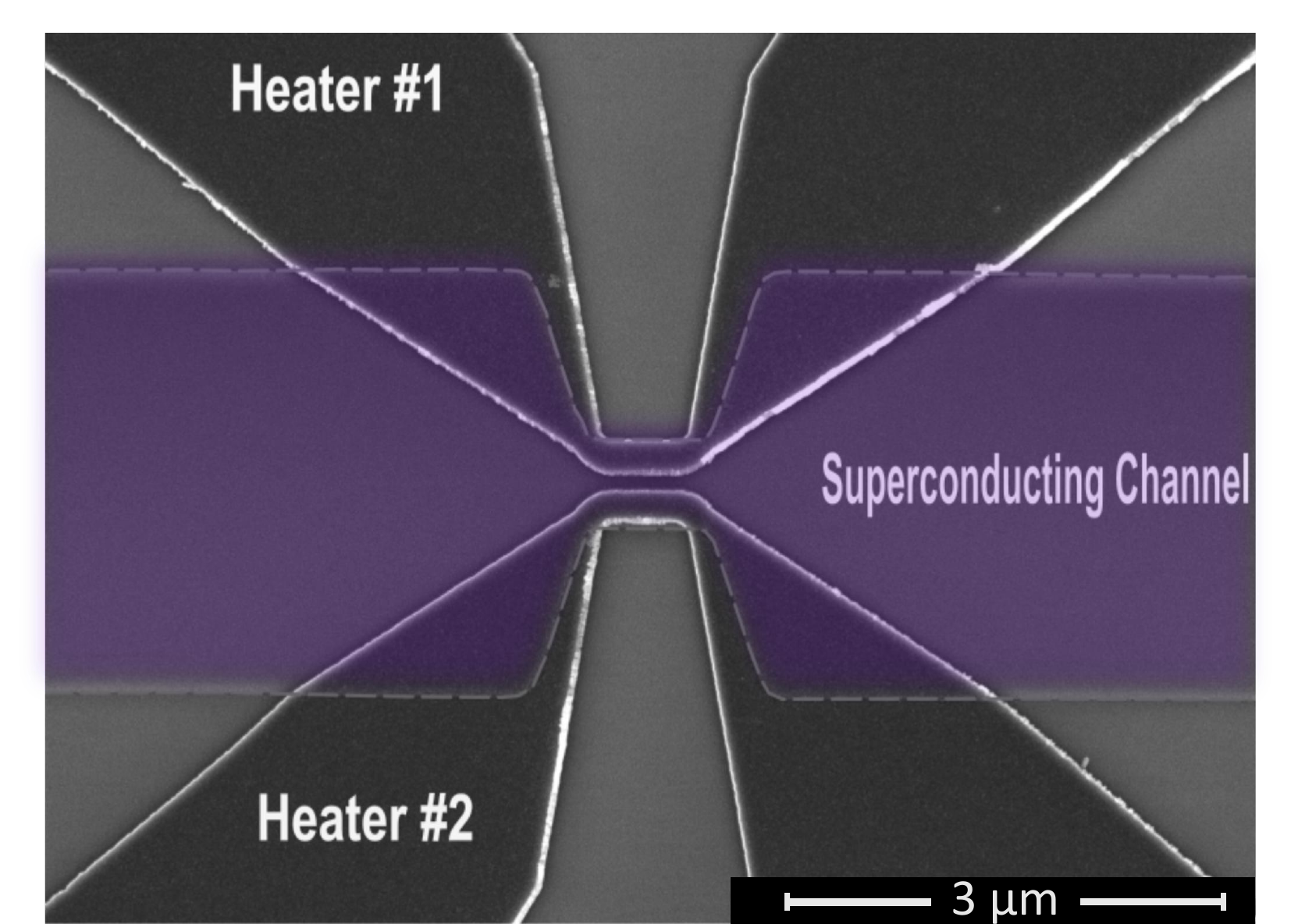}
        \caption{}
        \label{subfig:c}
    \end{subfigure}
    \caption{Dual-heater hTron (a) 3D Schematic (b) Scanning Electron Microscope (SEM) image of the fabricated sample.}
    \label{fig:main}
\end{figure}

As shown in the schematic in Fig.~\ref{fig:main}\subref{subfig:b10}, the proposed design features two metal heaters positioned parallel to the superconducting channel, separated by an insulating layer. In this configuration, both heaters contribute to the heat transfer to the superconducting channel and hence, independently or collectively (depending on the channel bias) can control the state of channel (superconducting or normal). 

The Scanning Electron Microscope (SEM) image of the fabricated chip is presented in Fig.~\ref{fig:main}\subref{subfig:c}. The heaters are physically separated from each other and also electrically isolated from the superconducting channel using a thin  ($\sim40\,\mathrm{nm}$) layer of Plasma Enhanced Chemical Vapor Deposition (PECVD) SiO$_2$
 layer. Heat is transferred via phonons to the superconducting channel, and once the local temperature exceeds the critical threshold (at a specific channel bias current), a resistive section is forms across the channel which disrupts superconductivity and diverts the current into the readout circuitry.

\begin{figure}[!htbp]
    \centering
         \begin{subfigure}{0.85\textwidth}
        \centering
        \includegraphics[width=\textwidth, keepaspectratio]{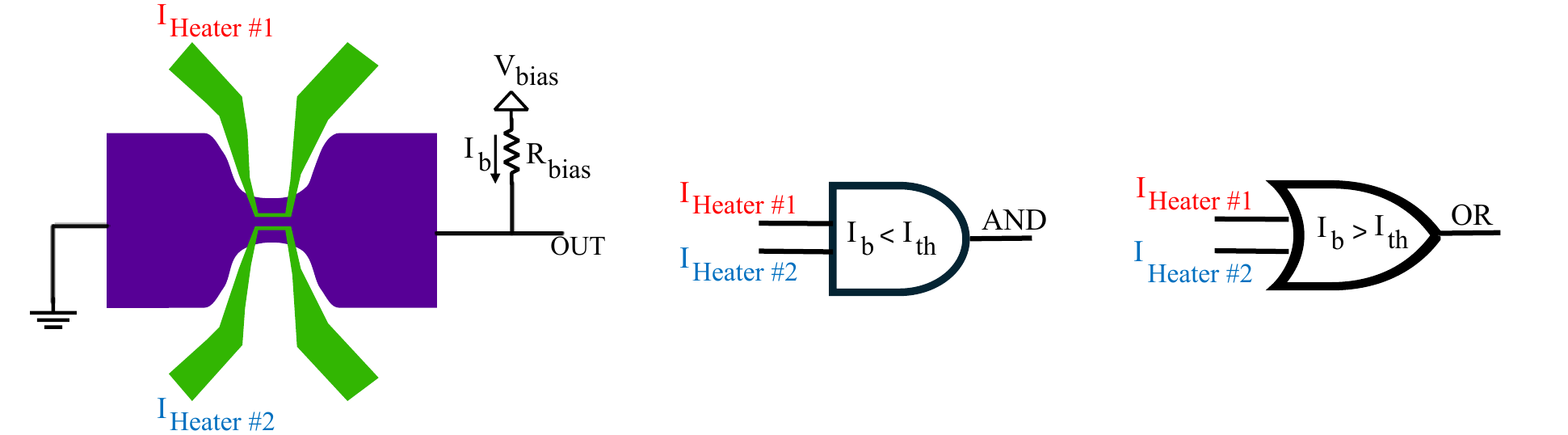}
        \caption{}
        \label{subfig:d1}
    \end{subfigure}
        \begin{subfigure}{0.85\textwidth}
        \centering
        \includegraphics[width=\textwidth, keepaspectratio]{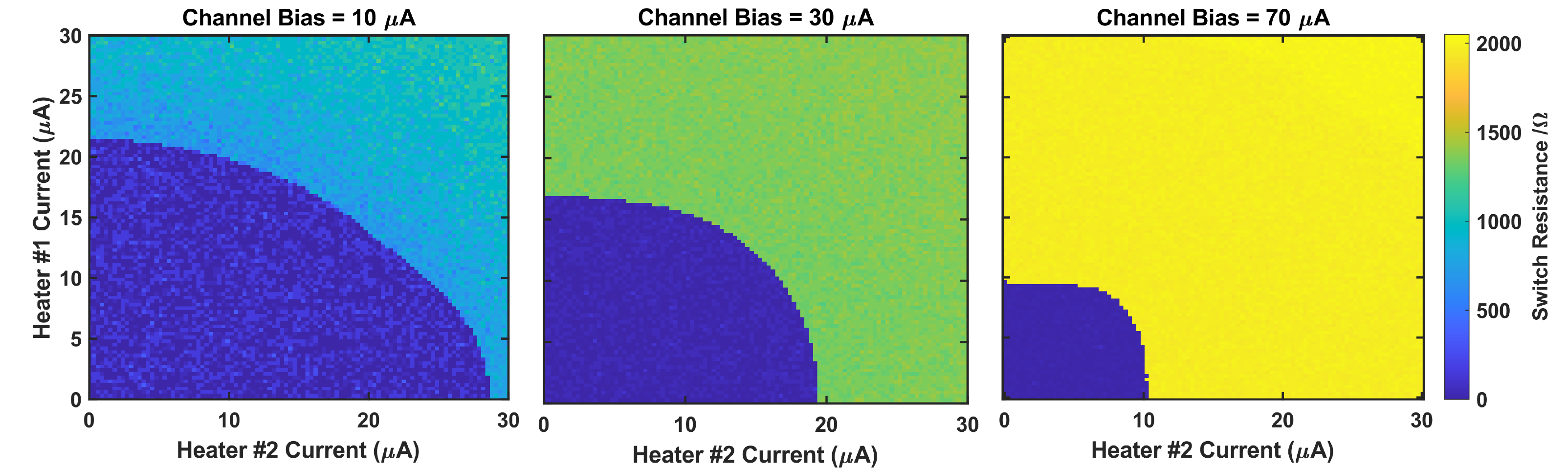}
        \caption{}
        \label{subfig:c1}
    \end{subfigure}
    \caption{DC characterization of the reconfigurable superconducting logic unit.
(a) Electrical measurement setup (left) showing the biasing configuration and output readout, together with the schematic diagram of the logic units.
(b) Measured superconducting channel resistance as a function of heater currents for three different channel bias values.
    }
    \label{fig:colormap}
    
\end{figure}

Figure~\ref{fig:colormap} illustrates the operating principle of the proposed superconducting logic unit. The electrical setup schematic used for DC characterization is shown in Fig. 2a. The device consists of a superconducting nanowire channel biased by a voltage source in series with a bias resistor, while two nano-heaters are independently driven by input currents. The output of the logic unit is defined as the voltage measured across the superconducting nanowire channel.

In the superconducting state, the nanowire channel exhibits zero electrical resistance; consequently, the voltage measured across the channel is zero. When thermal energy is supplied by the nano-heaters, the local temperature of the nanowire increases. Depending on the applied channel bias current and the amount of heat generated by the heaters, superconductivity may either be maintained or suppressed. Once superconductivity is locally disrupted, a resistive region can form within the nanowire, leading to a finite resistance and the appearance of a non-zero output voltage. Therefore, the measured output voltage directly reflects the state of the nanowire and serves as the logic output of the circuit.

For a fabricated chip, the device geometry remains fixed, with each nano-heater having a width of 70 nm and the superconducting nanowire having a width of 400 nm (for the devices shown in Figs. 2 and 3).
The resistance of both the nanowire and the heaters is determined by their dimensions and material properties. Therefore, the state of the nanowire is governed by the bias current applied to the superconducting channel together with the Joule heating generated by the nano-heaters. The contribution of each nano-heater, corresponding to each logic input, depends on its resistance and on the magnitude of the current flowing through it. The combined heating produced by the inputs determines the formation of thermal hotspots along the nanowire and ultimately controls the transition between the superconducting and resistive states.

Figure 2b presents experimental results obtained under steady-state DC conditions, showing the superconducting channel resistance as a function of the heater currents for three different fixed channel bias currents. The blue regions correspond to zero measured output voltage, indicating that the nanowire remains in the superconducting state. For lower bias currents, both heaters must be activated to generate sufficient hotspots and Joule heating to drive the nanowire into the resistive state, which reflects AND logic behavior.
As the channel bias current increases, a smaller heater current becomes sufficient to induce switching to the resistive regime, resulting in a reduction of the superconducting (blue) region. Under this condition, activation of either heater alone can trigger the transition, corresponding to OR logic behavior.

\begin{figure}[!htbp]
    \centering

    \begin{subfigure}{0.35\textwidth}
        \centering
        \includegraphics[width=\textwidth]{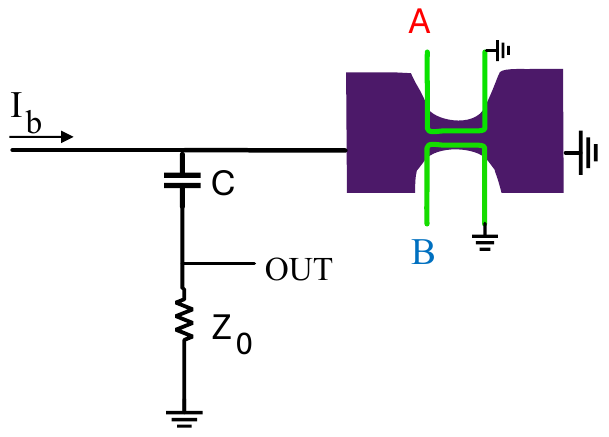}
        \caption{}
        \label{subfig:a2}
    \end{subfigure}
    \hfill
    \begin{subfigure}{0.5\textwidth}
        \centering
        \includegraphics[width=\textwidth]{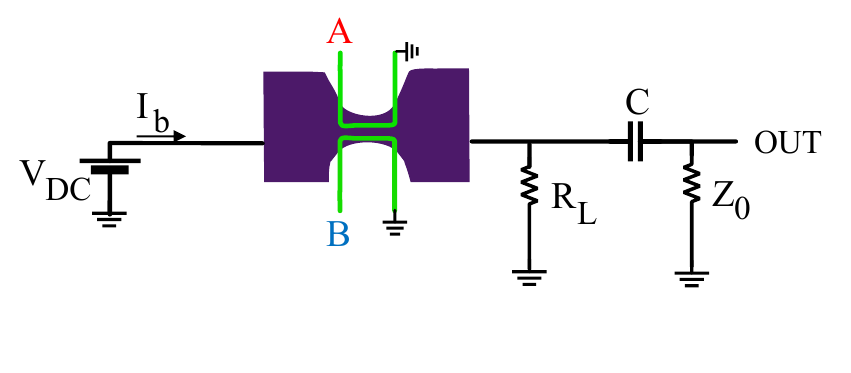}
        \caption{}
        \label{subfig:d2}
    \end{subfigure}


    \begin{subfigure}{0.45\textwidth}
        \centering
        \includegraphics[width=\textwidth]{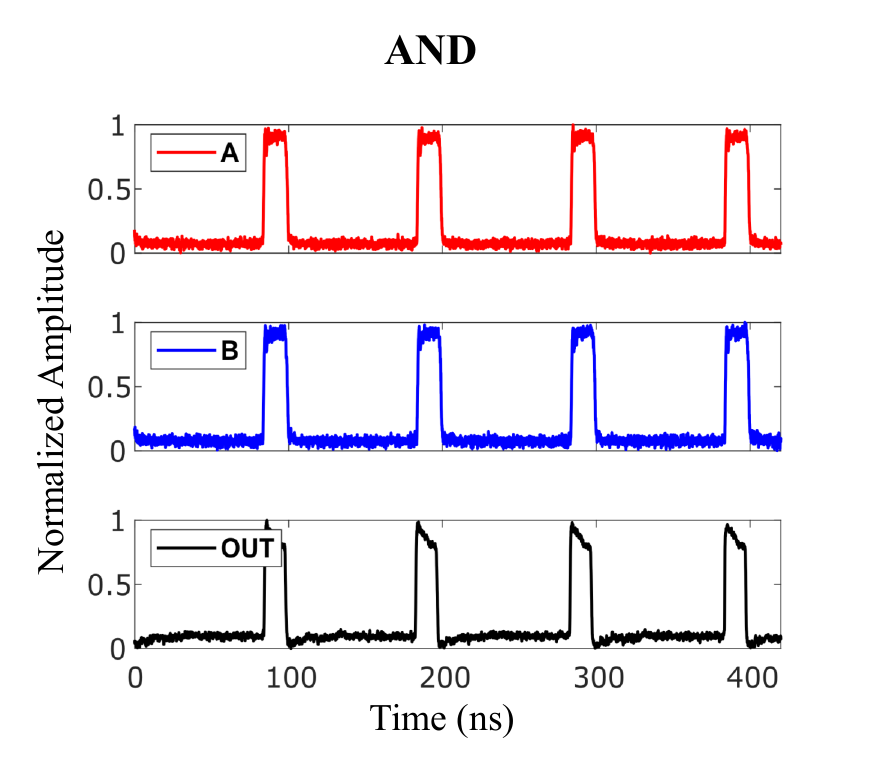}
        \caption{}
        \label{subfig:b2}
    \end{subfigure}
    \hfill
    \begin{subfigure}{0.45\textwidth}
        \centering
        \includegraphics[width=\textwidth]{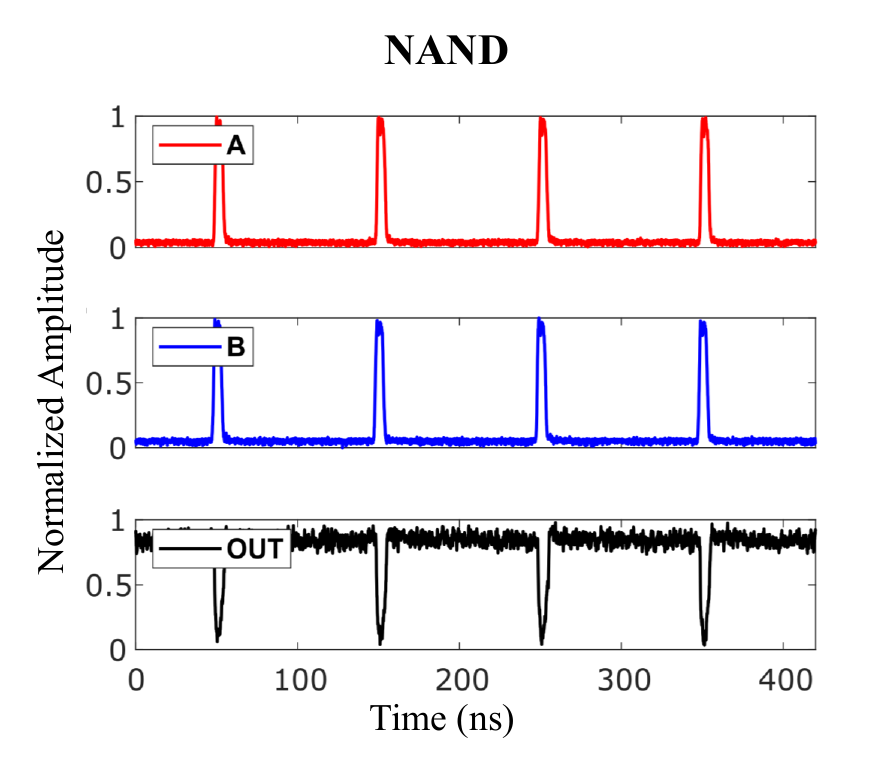}
        \caption{}
        \label{subfig:e2}
    \end{subfigure}


    \begin{subfigure}{0.45\textwidth}
        \centering
        \includegraphics[width=\textwidth]{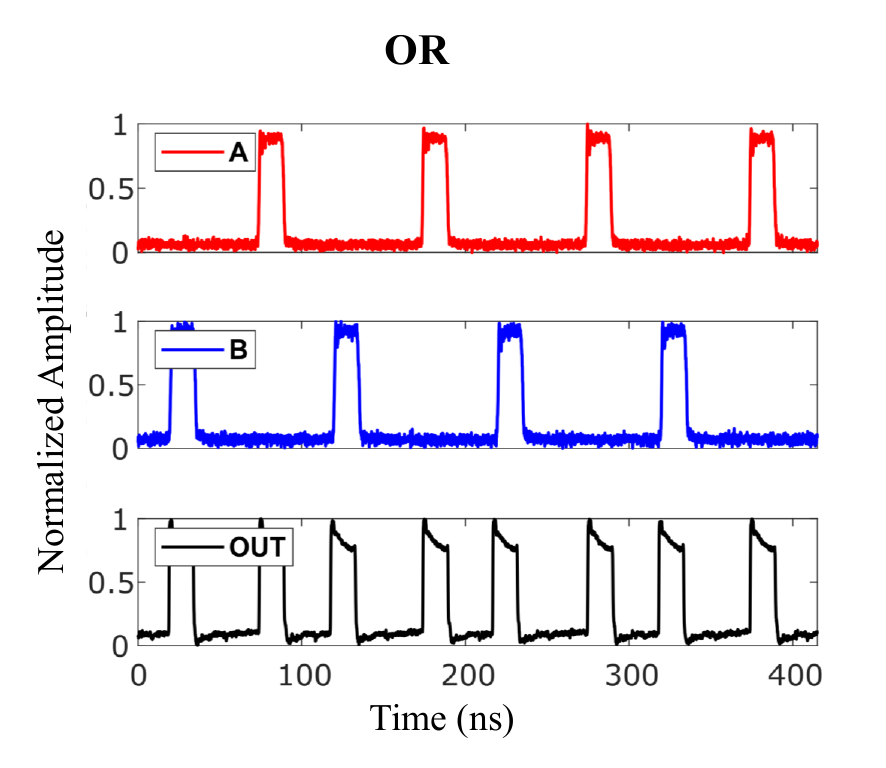}
        \caption{}
        \label{subfig:c2}
    \end{subfigure}
    \hfill
    \begin{subfigure}{0.45\textwidth}
        \centering
        \includegraphics[width=\textwidth]{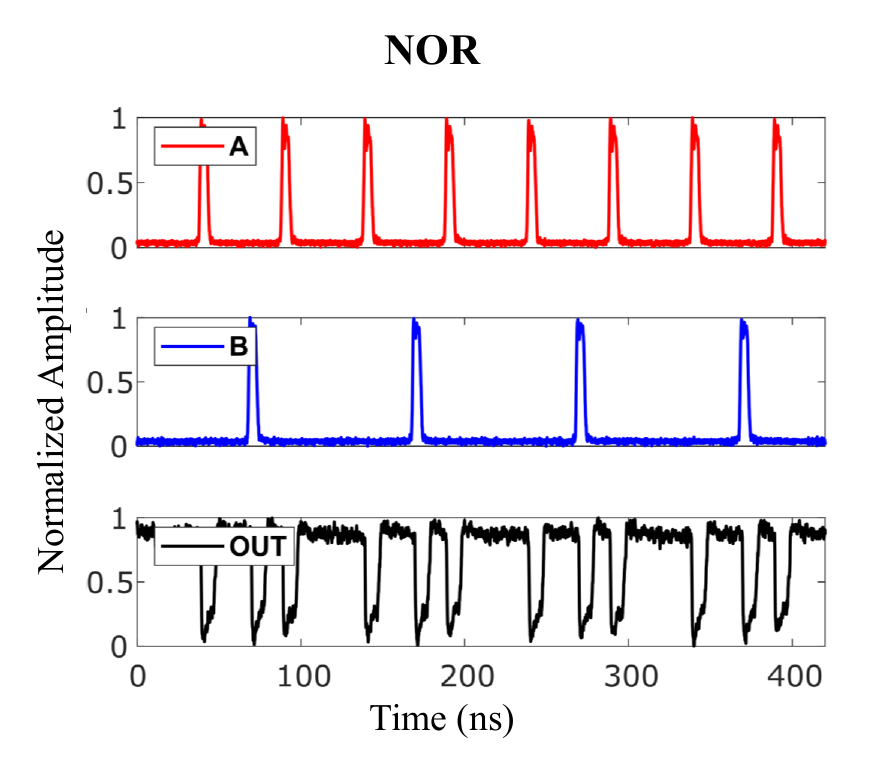}
        \caption{}
        \label{subfig:f2}
    \end{subfigure}

    \caption{Experimental realization of dual-heater hTron logic gates (a) measurement setup schematic for AND and OR logic gate (b) NAND/NOR measurement setup configuration (c) AND logic operation ($I_b = 35\,\mu\text{A}$) (d) NAND logic operation for DC voltage source set to 1 mV (e) OR logic function ($I_b = 55\,\mu\text{A}$) (f) NOR logic function ($V_{\mathrm{dc}}$=1.35 mV).}
    
    \label{fig:detection}
\end{figure}

AC characterization of the hTron device with two heaters further demonstrates its reconfigurable logic functionality. The schematic of the measurement setup for AND and OR operations is shown in Fig.~\ref{subfig:a2}. One contact pad of the superconducting channel is grounded, while the other serves as both the bias port and the output measurement point. The currents applied to the two nano-heaters act as logic inputs, and the resulting voltage across the nanowire defines the logic output.

The local heat generated by the heaters controls the superconducting state of the nanowire, and thus the output voltage. Detailed information about the experimental setup and measurement equipment is provided in the \textit{Supplementary material}.

For a fixed combination of heater input currents, the output voltage depends on the bias current applied to the superconducting channel. By adjusting this channel bias, the device can be dynamically configured to exhibit either AND or OR logic behavior. Specifically, when the bias is below a defined threshold, the nanowire remains superconducting unless both heaters are active, producing an AND response. When the bias exceeds this threshold, activation of either heater alone is sufficient to trigger a resistive transition, resulting in OR logic operation. Figures~\ref{fig:detection}\subref{subfig:b2} and ~\ref{fig:detection}\subref{subfig:c2}
show representative measurements of AND and OR behavior, respectively, demonstrating this reconfigurable operation.
In Fig.~\ref{fig:detection}\subref{subfig:b2} configuration, the superconducting nanowire transitions from the superconducting to the resistive state only when both heaters are activated simultaneously. When the channel current is $35\,\mu\text{A}$, the combined heat is necessary to generate stable hotspots that span the width of the nanowire, disrupting superconductivity. 
As presented in Fig.~\ref{fig:detection}\subref{subfig:c2}, in the OR configuration, the biasing current is sufficiently high such that even a small amount of heat generated by either heater is enough to drive the nanowire into a non-superconducting state. In other words, the activation of just one heater creates a hotspot that disrupts superconductivity, resulting in a measurable output voltage. 
Experimental results confirm consistent behavior with the expected OR logic operation, when the bias current is set to $55\,\mu\text{A}$.

The same structure can be adapted to generate additional logic operations by employing a different circuit configuration. In this setup, biasing is achieved by connecting one end of the nanowire to a voltage source.
The second end of nanowire is terminated by a $50\,\Omega$ shunt and a parallel capacitively coupled readout (Figure~\ref{fig:detection}\subref{subfig:d2}).
When the nanowire is in the superconducting state, its resistance is effectively zero, hence the source voltage can be measured at the output.
In contrast, when superconductivity is disrupted, the nanowire exhibits a high resistance, and the voltage drops across the nanowire itself and a significantly smaller voltage, corresponding to a logic low, is measured at the output.
This transition allows for clear detection of state changes and facilitates the implementation of different logic functions. 

By adjusting the current flowing through the channel to remain below a threshold value, the system exhibits behavior corresponding to a NAND logic operation. 
When none or only one of the heaters is activated, the nanowire remains in the superconducting state, and the output voltage stays at a high level. However, when both heaters are activated simultaneously, the combined heat is sufficient to drive the nanowire into the resistive state, resulting in a drop in output voltage to a low level. 
The experimental results, including the input signals applied through the heaters and the resulting output voltage, are presented in Fig.~\ref{fig:detection}\subref{subfig:e2}. 

The circuit configuration and cabling remain unchanged from the previous section; only the input voltage is adjusted to realize NOR logic behavior. At the selected bias current, heating from a single heater is sufficient to switch the superconducting nanowire to the resistive state. Consequently, the activation of either input (or both) produces a low output voltage, while the absence of both inputs maintains a high output voltage (Fig.~\ref{fig:detection}\subref{subfig:f2}). The measured results agree well with the expected physical behavior, confirming NOR logic operation.
The measured output voltage aligns well with the expected physical behavior and supports the implementation of NOR logic. 
\begin{figure}[htbp]
    \centering

    \begin{subfigure}{0.5\textwidth}
        \centering
        \includegraphics[width=\linewidth]{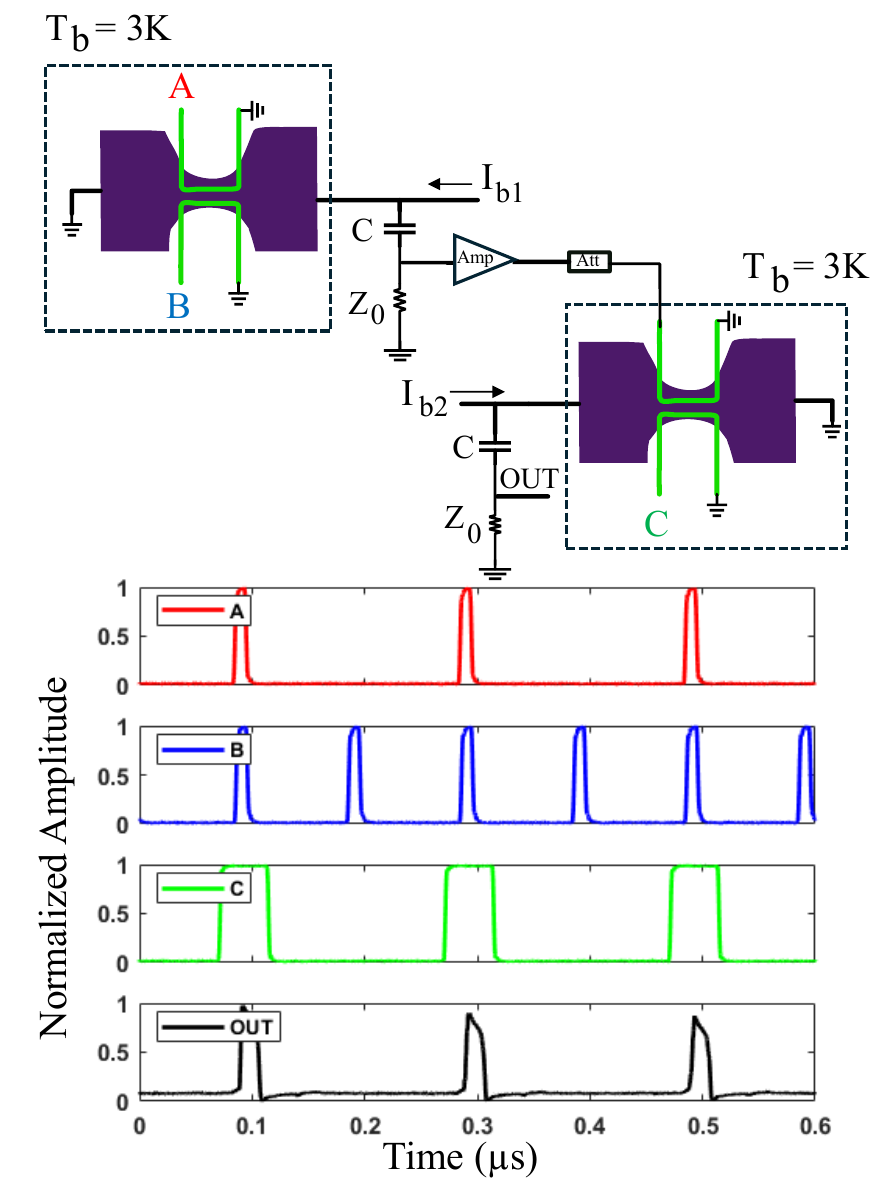}
        \caption{}
        \label{subfig:a}
    \end{subfigure}
    \hfill
    \begin{subfigure}{0.49\textwidth}
        \centering
        \includegraphics[width=\linewidth]{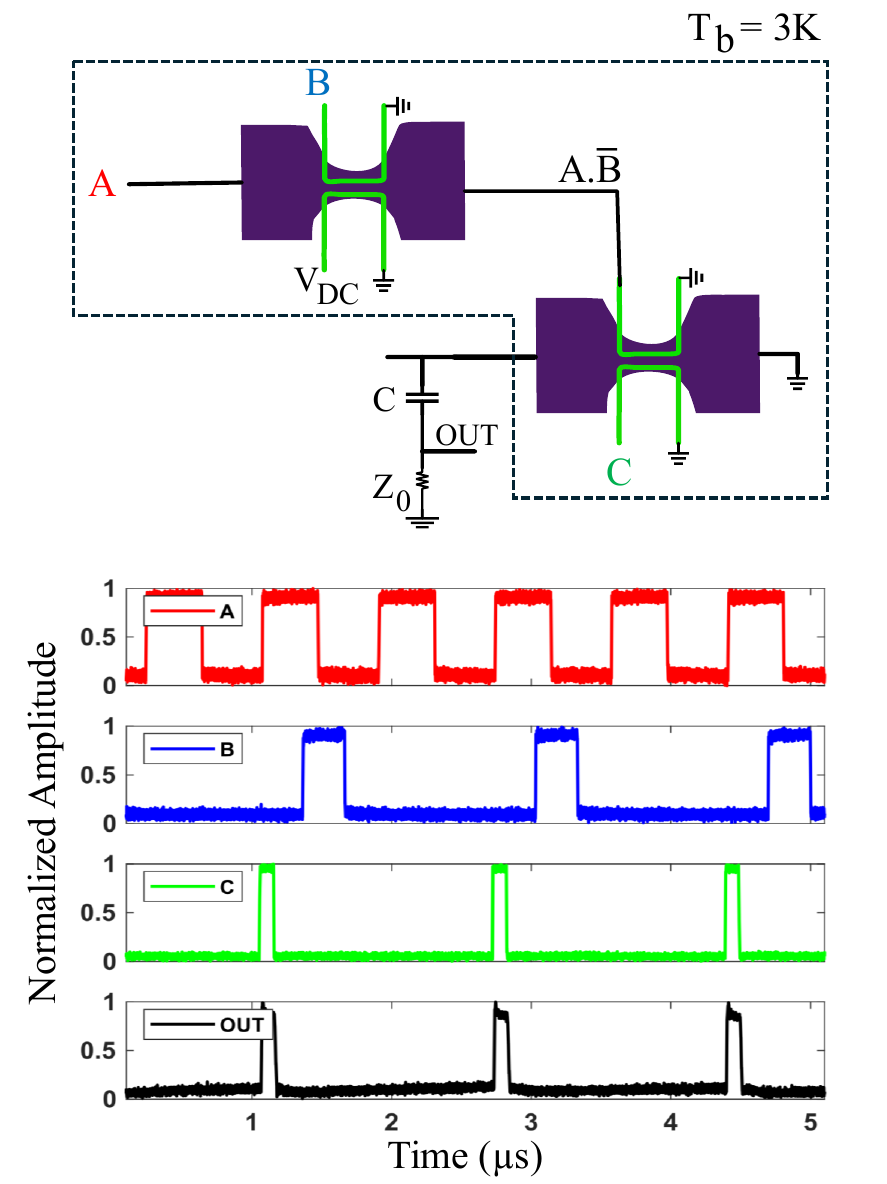}
        \caption{}
        \label{subfig:b}
    \end{subfigure}

    \caption{ Interconnecting two hTrons
    (a) Off-chip configuration in which one hTron is driven by the output of another hTron (top), along with the measured input and output signals representing $(A \cdot B \cdot C)$ logic.
(b) Schematic of the on-chip connection between two hTron devices (top) and experimental data demonstrating the implementation of $(A \cdot \overline{B} \cdot C)$.
    }
    \label{fig:Integration}
\end{figure}

So far, the structure introduced allows for implementation of different logic functionalities through adjustments of biasing and circuit configuration. Building upon this, we explore the scalability of our platform by cascading two unit-cells to implement larger logics. This integration opens the door to utilizing these switches as functional elements for universal programmable logics. 
In this context, hTron can be arranged in layers, where each layer is responsible for a level of computation. The input to each layer is derived from the output of the preceding layer, enabling signal propagation and complex decision-making processes like artificial neural networks. 
As an initial step toward integration, we investigated the simplest configuration: taking the output of one hTron and using it as the input to another. This foundational experiment serves as a proof of concept for cascading operations, forming the basis for more advanced circuit designs. 
As a first step toward switching a dual-input hTron using the output of another one, the electrical connection is implemented outside the cryostat (Figure ~\ref{fig:Integration}\subref{subfig:a}). In this configuration, the output signal is collected using room-temperature instrumentation. The monitored signal is amplified and subsequently attenuated to make it suitable as the input for the next dual-input hTron. This conditioned signal is then applied to the heaters, providing the required input to trigger the second switch; details of the instrumentation and parameters are provided in the \textit{Supplementary material}.

For this demonstration, both devices employ nanowire channels with a width of 1.3~µm. Within each hTron, the two heaters have identical geometry, while the heater dimensions differ between the first and second hTron (with widths of 300~nm and 200~nm, respectively).
Figure~\ref{fig:Integration}\subref{subfig:a} shows the circuit schematic used for the off-chip integration of a dual-heater hTron. The DC bias current ($I_b < I_{th}$) of the superconducting channel is selected to implement AND logic for each hTron. If the two combined layers are considered as a single gate, the resulting output voltage reflects the combined effect of the consecutive operations. As a result, a three-input AND function can be realized $(A \cdot B \cdot C)$. Therefore, the output signal from the first hTron is required to drive the second stage, enabling a state change and the generation of an output signal. The input pulse signals applied to the heaters and the corresponding output are shown in bottom plot Fig.~\ref{fig:Integration}\subref{subfig:a}. We can make use of the reconfigurability properties at the second stage and set the bias current to obtain OR logic, as reported in the \textit{Supplementary material}. 

To fully demonstrate and showcase the capabilities of the device, we implement a fully integrated configuration.
 In Figure ~\ref{fig:Integration}\subref{subfig:b} setup, no external amplifier or attenuator was used. All processing was performed at cryogenic temperatures, indicating that all interconnections were implemented on-chip within the cryostat. The corresponding input signals and output voltages are illustrated in Figure ~\ref{fig:Integration}\subref{subfig:b} (bottom). Due to loading by the measurement equipment, we only accessed to the final output signal. The behavior of the first dual-heater hTron is controlled by their inputs, and the input to the second stage is adjusted to ensure it could be reliably triggered by the output of the first stage. 
 
This configuration effectively results in more complex operations, where the output of the first stage $(A \cdot \overline{B})$ serves as the input for the second stage. For the second-stage hTron, which is driven by the output of the preceding device, the channel bias current determines the logic function it can perform. In this case, the device is configured to operate as an AND gate, and the corresponding output signal is shown in the measured graph, representing $(A \cdot \overline{B} \cdot C)$. The Boolean operation $(A \cdot \overline{B} + C)$ can also be implemented, as described in the \textit{supplementary material}.
This approach paves the way for integrating multiple hTrons, facilitating the realization of complex digital functionalities on a cryogenic platform. Such integration allows, for example, a raw data from superconducting nanowire single photon detectors (SNSPDs) array to be processed directly within the same environment. 

In conclusion, we have introduced a new reconfigurable logic structure that operates based on the principles of superconductivity. Two heaters are used to generate Joule heating and create a hotspot across a nanowire. The state of this dual-input hTron depends on the bias current applied through the nanostructure and the heaters. 
What makes this structure novel is its ability to perform multiple logic operations in the same device. By adjusting the currents flowing through the inputs, the device can be configured to function as either an AND or OR gate, or as a NAND or NOR gate, depending on the electrical setup. This unique behavior eliminates the need for multiple separate logic elements for each operation, thereby reducing area and design complexity. The proposed device can perform NAND and NOR (universal logic gates) 
giving a complete set of logic functions that can be used to build any Boolean circuit.
Furthermore, we demonstrate that the output of a dual-heater hTron can drive another one, allowing for higher-level integration and computational functionality. This suggests that hTrons could potentially serve as building blocks in superconducting circuits proposed for computing architectures inspired by neural networks \cite{Hidaka1993_SuperconductingNeuralCell, Goteti2021_SuperconductingNN}, which aim for energy-efficient and low-noise processing.
\section*{Supplementary Material}
See the supplementary material for details on sample fabrication, electrical measurement setup, and DC characterization of the dual-input hTrons. The dynamic switching energy required to operate the device is evaluated as a function of operating frequency. Furthermore, cascaded on-chip operation is demonstrated, highlighting the use of device reconfigurability to realize additional logic functionalities beyond those presented in the main text.
\begin{acknowledgments}
I.~E.~Z. and B. B. acknowledge funding from the European Union’s Horizon Europe research and innovation programme under grant agreement No. 101098717 (RESPITE project) and No.101099291 (fastMOT project); I. E. Z. acknowledges the funding from the FREE project (P19-13) of the TTW-Perspectief research program partially financed by the Dutch Research Council (NWO).
\end{acknowledgments}
\section*{Author Declarations}
\subsection*{Conflict of Interest}
M. Mikhailov (full-time), and I. Esmaeil Zadeh (part-time) are employed by Single Quantum B.V. and may profit financially. The authors have no conflicts to disclose.

\subsection*{Author Contributions}
\textbf{B. Babaghorbani}: Conceptualization (equal); Data curation (equal); Formal analysis (equal); Investigation (lead); Methodology (equal); Resources (equal); Software (equal); Supervision (equal); Validation (equal); Visualization (equal); Writing – original draft (equal); Writing – review \& editing (equal). \textbf{M. Mikhailov}: Investigation (equal); Methodology (equal); Resources (equal); Writing – original draft (equal); Writing – review \& editing (equal). \textbf{H. Wang}: Data curation (equal); Formal analysis (equal); Investigation (equal); Writing – original draft (equal); Writing – review \& editing (equal). \textbf{T. Descamps}: Investigation (equal); Resources (equal); Writing – original draft (equal); Writing – review \& editing (equal). \textbf{V. Zwiller}: Supervision (equal); Writing – original draft (equal); Writing – review \& editing (equal). \textbf{I. Esmaeil Zadeh}: Conceptualization (equal); Funding acquisition (equal); Investigation (equal); Methodology (equal); Project administration (equal); Supervision (equal); Validation (equal); Visualization (equal); Writing – original draft (equal); Writing – review \& editing (equal).
\section*{Data Availability}
The data that support the findings of this study are available on request from the authors.

\bibliography{reconfigurable_ref}

\end{document}